\documentclass[prl,twocolumn,showpacs,twoside,superscriptaddress]{revtex4}
\usepackage{graphicx}
\usepackage{amsmath}
\usepackage{placeins}
\usepackage{hyperref}
\usepackage{bbm}

\newcommand{\eq}{\begin{eqnarray}}
\newcommand{\en}{\end{eqnarray}}

\newcommand{\eqq}{\refstepcounter{equation} \begin{equation} \tag{\roman{equation}} }
\newcommand{\enn}{\end{equation}}

\def\ket#1{\mathinner{|{#1}\rangle}}
\newcommand{\braket}[2]{\langle #1|#2\rangle}

\begin{document}

\title{Collective interference of composite two-fermion bosons} 

\author{Malte C. Tichy}
\affiliation{Lundbeck Foundation Theoretical Center for Quantum System Research,
Department of Physics and Astronomy, University of Aarhus, DK-8000 Aarhus C, Denmark}

\author{Peter Alexander Bouvrie}
\affiliation{Departamento de F\'isica At\'omica, Molecular y Nuclear and Instituto Carlos I de F\'isica Te\'orica y Computacional, Universidad de Granada, E-18071 Granada, Spain}

\author{Klaus M\o{}lmer}
\affiliation{Lundbeck Foundation Theoretical Center for Quantum System Research,
Department of Physics and Astronomy, University of Aarhus, DK-8000 Aarhus C, Denmark}

\date{\today}


\begin{abstract}
The composite character of two-fermion bosons manifests itself in the interference of many composites as a deviation from the ideal bosonic behavior. A state of many composite bosons can be represented as a superposition of different numbers of perfect bosons and fermions, which allows us to provide the full Hong-Ou-Mandel-like counting statistics of interfering composites. Our theory quantitatively relates the deviation from the ideal bosonic interference pattern to the entanglement of the fermions within a single composite boson. 
\end{abstract}

\maketitle

The quantum statistics of bosons is most apparent in correlation functions and counting statistics. Characteristic bosonic signatures are  encountered for thermal states, which feature the Hanbury Brown and Twiss effect \cite{BROWN1956,Jeltes2007,Molmer2008,Perrin2010}, as well as in meticulously prepared Fock-states \cite{Hong1987,Lim2005,Laloe2011,tichyunpu}, which exhibit Hong-Ou-Mandel-like  (HOM)  interference. Deviations from the ideal bosonic pattern in HOM setups are often caused by inaccuracies in the preparation of Fock-states and in the alignment of the setup, which induce partial distinguishability between the particles \cite{Hong1987,Ou1999,Tichy2011,Ra2011}. Another source for deviations from perfect bosonic behavior has received only little attention, limited to mixed states \cite{Combescot,Ramanathan2011a}: Since most bosons are composites (``cobosons'') made of an even number of fermions, reminiscences of underlying fermionic behavior are expected in many-coboson interference. In analogy to partially distinguishable particles \cite{Tichy2011,Ra2011}, one can intuitively anticipate that the many-coboson wave-function partially behaves in a fermionic way, with impact on the resulting counting statistics. 

Here, we investigate such compositeness effects in HOM interferometry of cobosons. The ideal bosonic interference pattern is jeopardized by the Pauli principle that acts on the underlying fermions, an effect that becomes relevant when the constituents populate only a small set of single-fermion states. 
 The effective number of single-fermion states can be related to the entanglement between the fermions, via the Schmidt decomposition. 
  Not only does entanglement thus   guarantee the irrelevance of the Pauli-principle for coboson states, but it also constitutes the very many-body coherence property that ensures that many-coboson interference matches the ideal bosonic pattern \cite{Laloe2011,tichyunpu}. 
   The many-coboson wavefunction can be described as a superposition of different numbers of perfect bosons and fermions, with weights that are determined by the Schmidt coefficients. Using that intuitive representation, we compute the exact counting statistics in many-coboson interference and thus provide direct experimental observables for compositeness. 
Properties of the collective wave-function of the fermionic constituents can thus be extracted from coboson interference signals, while in the limit of truly many particles, particularly simple forms for the interference pattern emerge.

The bottomline of our discussion, the observable competition of fermions for single-particle states, is a rather general phenomenon that is not restricted to any particular physical system. To render our analysis of many-coboson interference tangible, however, we focus on an interferometric setup that can be realized with trapped ultracold atoms \cite{supplementary}. 

\begin{figure}[th]
\center
\includegraphics[width=.8\linewidth,angle=0]{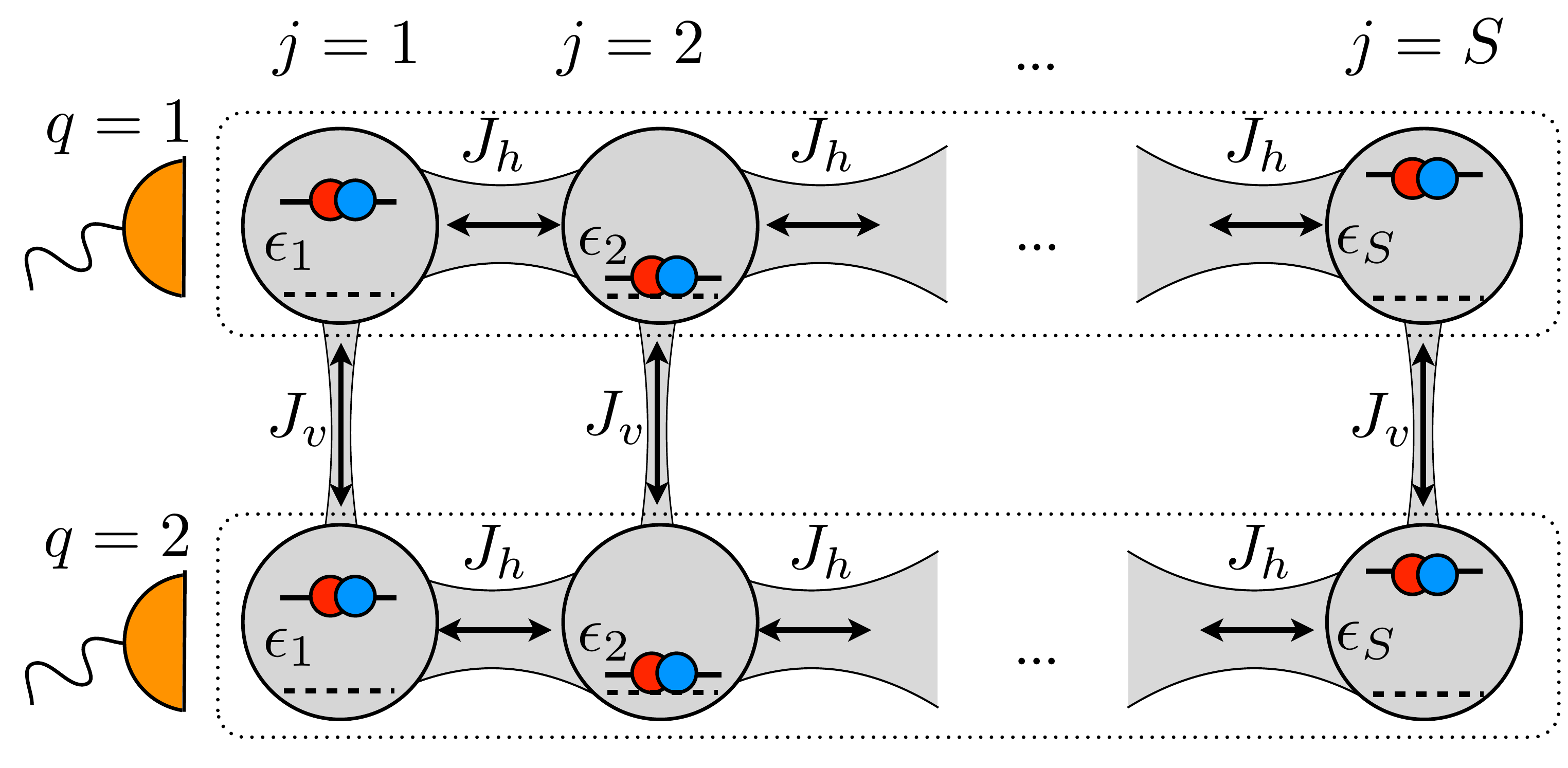} 
\caption{Setup for the interference of engineered cobosons. $N_1$ ($N_2$) strongly bound bi-fermions are prepared in the upper (lower) lattice at ${J_v} \ll {J_h} $, such that each bi-fermion is governed by the local energies $\epsilon_j$ and the tunneling rate ${J_h}$. The barrier between the lattices is then ramped down, such that ${J_v} \gg {J_h}$  and vertical tunnelling takes place. The total number of bi-fermions in the upper and lower lattice is then counted.} 
 \label{scheme2}
\end{figure}

We consider strongly bound bi-fermion pairs that are trapped in a two-dimensional potential landscape with different horizontal and vertical coupling rates \cite{Strzys2010}, as depicted in Fig.~\ref{scheme2}, which is described by the Hamiltonian
\eq 
\hat H & = &  \nonumber - \frac {J_h} 2  \sum_{q=1}^2 \sum_{j=1}^{S-1}  \hat d^\dagger_{q,j} \hat d^{\phantom \dagger}_{q,j+1}   
 - \frac {J_v} 2 \sum_{j=1}^S  \hat d^\dagger_{1,j} \hat d^{\phantom \dagger}_{2,j}   + h.c. \\ 
&& +\sum_{q=1}^2 \sum_{j=1}^S  \epsilon_j \left( \hat d^\dagger_{q,j} \hat d_{q,j}^{\phantom \dagger} \right) , \label{hamil}
\en
where 
$\hat d_{q,j}^\dagger= \hat a_{q,j}^\dagger \hat b_{q,j}^\dagger$ creates a bi-fermion consisting of an $a$- and a $b$-type fermion in the $j$th site of the upper or lower lattice ($q=1,2$); ${J_h}$ (${J_v}$) is the effective tunneling strength along (between) the lattices, and $\epsilon_j$ defines a local energy landscape \cite{supplementary}. We assume that, initially, ${J_h} \gg {J_v}$, and multi-coboson states are prepared in the horizontally extended lattice $q$ by \cite{Law2005,Chudzicki2010} 
\eq 
\hat c_{q}^\dagger = \sum_{j=1}^S \sqrt{\lambda_{j}} ~ \hat d_{q,j}^\dagger = \sum_{j=1}^S \sqrt{\lambda_{j}} ~ \hat a_{q,j}^\dagger \hat b_{q,j}^\dagger  \label{compositeboson} .
\en
A coboson is thus a horizontally delocalized bi-fermion, and the $S$  coefficients $\lambda_j$ are then the \emph{Schmidt coefficients} of the two-fermion state.  

The distribution $\vec \lambda$ is conveniently characterized by its moments \eq M(m) = \sum_{j=1}^S \lambda_j^m ,\en
where normalization implies $M(1)=1$ and $M(2)=P$ is the purity of either reduced single-fermion state. 
We consider an initial state of $N_1$ cobosons in the upper and $N_2$ cobosons in the lower lattice \cite{supplementary}, 
\eq 
\ket{\Psi} =  \frac{\left( \hat c_1^\dagger \right)^{N_1} }{\sqrt{ \chi_{N_1}  \cdot N_1!}} \frac{\left( \hat c_2^\dagger \right)^{N_2} }{ \sqrt{ \chi_{N_2} \cdot N_2!}} \ket{0} , \label{inistate}
\en
where we assume $N_1 \ge N_2$, and $ \chi_N$
is the coboson normalization factor \cite{Combescot2003,Law2005,Chudzicki2010,Combescot2008a,ourselves}, a symmetric polynomial 
\cite{Macmahon1915} given by $\chi_N=\Omega(\{ \underbrace{1, \dots ,1}_N \} ) $, with 
\eq 
\Omega(\{x_1, \dots, x_N \} ) &=& \sum_{\substack{ p_1, \dots , p_N \\  1\le p_j \le S }}^{i \neq j \Rightarrow p_i \neq p_j} \prod_{q=1}^N \lambda_{p_q}^{x_q} .
\en

To assess the behavior of the cobosons, we let the bi-fermions tunnel vertically between the two lattices by setting $J_v \gg J_h$ and  letting the system evolve for a time of the order $1/J_v$. Thus, beam-splitter-like dynamics couples the two lattices, while tunneling processes within the lattices, induced by $J_h$, can be neglected on this time-scale. The Schmidt modes $j$ are therefore left unchanged. 
 Time-evolution until $t$ implements a beam-splitter with reflectivity $R=\cos^2 \left(t {{J_v}} / {2} \right)$. 
In principle, the counting statistics of bi-fermions in the two lattices can be obtained by integrating the dynamics induced by (\ref{hamil}) for the initial state $\ket{\Psi}$ given in Eq.~(\ref{inistate}) and taking the expectation values of the counting operators 
\eq 
\hat A_{n_1,n_2} = \sum_{ \substack{   j_1 \neq j_2 \neq  \dots \neq  j_{n_1}, \\ l_1 \neq  l_2 \neq \dots \neq l_{n_2}} }^{ 1 \le j_k, l_m \le S  }
\prod_{k=1}^{n_1} \hat d^\dagger_{1,j_{k}}  \hat d^{\phantom \dagger}_{1,j_{k}}   \prod_{m=1}^{n_2} \hat d^\dagger_{2,l_{m}}  \hat d^{\phantom \dagger}_{2,l_{m}}    ,
\en
which witness the probability to find exactly $n_1$ ($n_2$) bi-fermions in the first (second) lattice. 
  This procedure, however, is computationally expensive and does not offer an intuitive physical picture. By exploiting the symmetry properties of the state (\ref{inistate}), one can show \cite{supplementary} that the behavior of cobosons is imitated exactly by a superposition of states with a different number of perfect bosons and fermions, in analogy to partially distinguishable particles \cite{Tichy2011,Ra2011}. When the distribution of the bi-fermions along the lattices is neglected, $\ket{\Psi}$ exhibits the same total counting statistics in the two lattices as the state
\eq 
\ket{\psi}&=& \sum_{p=0}^{N_2} \sqrt{w_p} \ket{\phi(p)} , \text{     with} \label{intuitivedecomposition} \\
\ket{\phi(p)} &=&
 \left[ \prod_{q=1}^2  \frac{\left( \hat g_q^\dagger \right)^{N_q-p} }{\sqrt{(N_q-p)! }} \right]
 \left[ \prod_{j=1}^p \hat f_{1,j}^\dagger \hat f_{2,j}^\dagger \right] \ket{0} , \label{Phipstate}
\en
 where $\hat g_{q}^\dagger$ ($\hat f_{q,j}^\dagger$) creates a boson ($j$-type fermion) in the lattice $q$. The weight of the component with $p$ pairs of fermionically behaving bi-fermions depends on the Schmidt coefficients and reads \cite{supplementary} 
\eq 
\label{weight}
w_p=  {N_1 \choose p }  {N_2 \choose p } ~\frac{p!}{\chi_{N_1} \chi_{N_2} } ~\Omega(\{ \underbrace{2,\dots, 2}_p,\underbrace{1,\dots ,1}_{N_1+N_2-2p}\}) .
\en

Combinatorially speaking, $w_p$ is the probability that, given two groups of $N_1$ and $N_2$ objects with properties distributed according to $\vec \lambda$, and assuming that all objects in either group carry different properties, one finds $p$ pairs of objects with the same property when the two groups are merged. In the present context, $w_p$ denotes the population of the state components in which the Pauli principle affects $p$ pairs of bi-fermions.  The term $\ket{\phi(0)}$ thus describes perfect bosonic behavior, its weight 
$ w_0=\chi^{\phantom \dagger}_{N_1+N_2}/ ({\chi^{\phantom \dagger}_{N_1} \chi^{\phantom \dagger}_{N_2}} )$  
 can be bound via the purity $P$ and the particle numbers $N_1,N_2$ \cite{Chudzicki2010,ourselves}: 
\eq 
\frac{  ( L-N_1)!   ( L-N_2)! }{ (L - N_1-N_2)! L! }  \le  w_0  \hspace{2.985cm} \label{boundsw0} \\ \nonumber \le 
 \frac{(1 - \sqrt P) (1 + \sqrt P  (N_1+N_2 - 1)) }{(1 + \sqrt P (N_2 - 1))(1 + \sqrt P (N_1 - 1))} ,
\en
where $L=\left \lceil \frac 1 P  \right \rceil $.

We can now derive the counting statistics of cobosons after time-evolution until $t=\pi/2/{J_v}$, which corresponds to a balanced beam-splitter with $R=T=1/2$. The probability $P_{\text{tot}}(m)$ to find $m$ cobosons in the upper lattice is the sum of the resulting probabilities from the different contributions in (\ref{intuitivedecomposition}), 
\eq 
P_{\text{tot}}(m)=\sum_{p=0}^{N_2} w_p \cdot P(m,p) , \label{weightedsumprobs}
\en
where $P(m,p)$ is the probability to find $m$ particles of any species in the upper lattice, given the state $\ket{\phi(p)}$ defined in (\ref{Phipstate}) and the beam-splitter  reflectivity $R=1/2$  \cite{supplementary}. 

\begin{figure}[th]
\center
\includegraphics[width=\linewidth,angle=0]{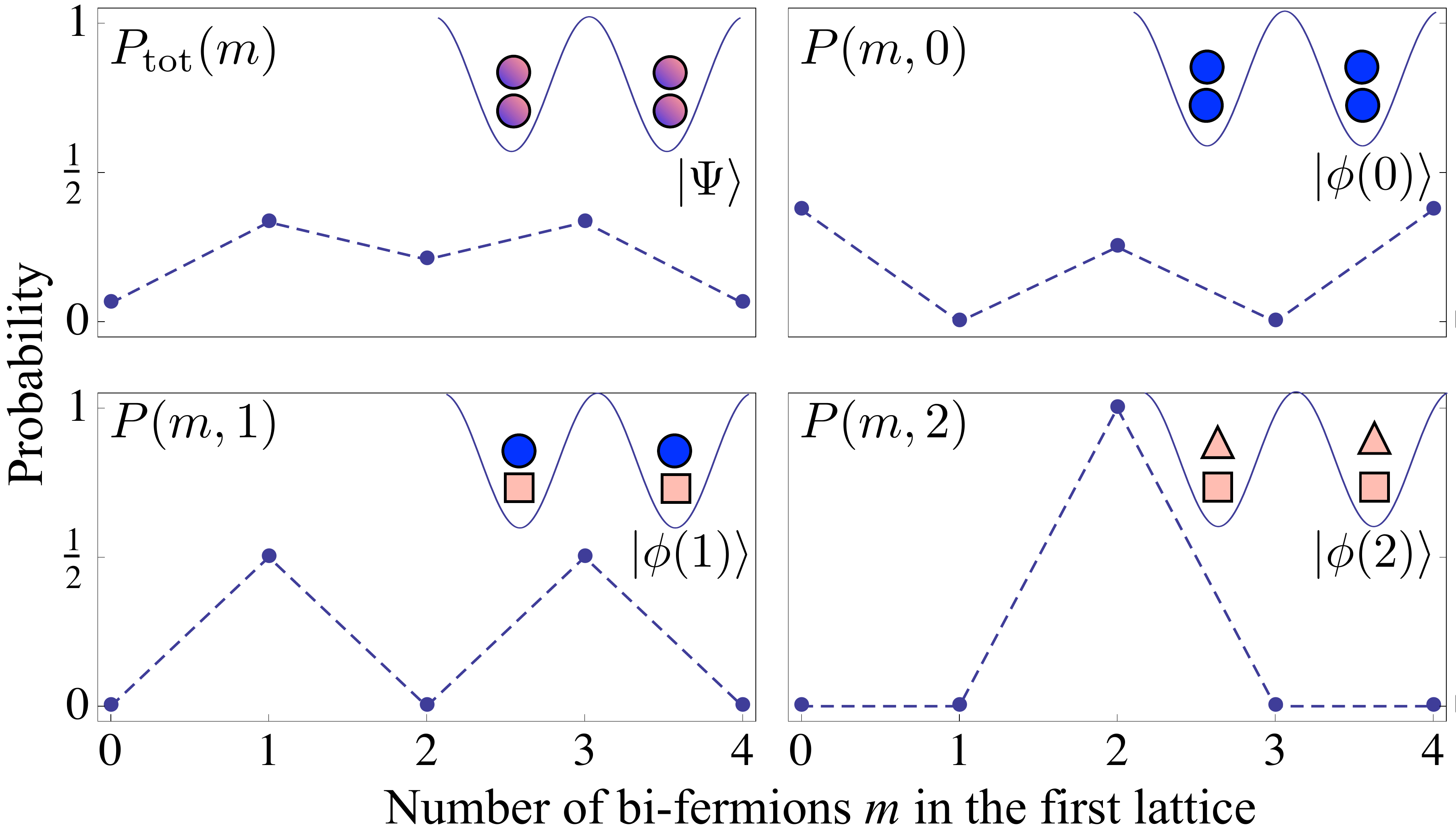} 
\caption{(color online) Counting statistics for the coboson-state $\ket{\Psi}$ with $N_1=N_2=2$, and of its components with different numbers of bosons and fermions $\ket{\phi(p)}$, ${p=0,1,2}$. Dark blue circles represent bosonically behaving bi-fermions, light orange symbols stand for fermionically behaving bi-fermions. The total counting statistics $P_\text{tot}(m)$ is the weighted sum (\ref{weightedsumprobs}) over the different components of the wave-function. While $\ket{\phi(0)}$ exhibits perfect bosonic behavior, $\ket{\phi(p \ge 1)}$ are partially fermionic, which leaves a signature in the counting statistics. Here, $R=1/2$ and $\lambda_1=\dots =\lambda_4=1/4$, such that $w_0=w_2=1/6$, $w_1=2/3$.} 
 \label{decompositions}
\end{figure}

 The simplest case is given by two interfering cobosons ($N_1=N_2=1$), for which we find $w_0=P$ and $w_1=1-P$: 
\eq 
P_{\text{tot}}(1)= P, \ \ P_{\text{tot}}(0)= P_{\text{tot}}(2) =  \frac{1-P}{2} . \label{HOM2signal}
\en

For ${P \rightarrow 1}$, the Pauli principle dominates and one always finds one particle in each lattice. In contrast to the interference of unbound boson pairs that can break up dynamically  \cite{Brougham2010}, a perfect bosonic dip emerges here in the limit of vanishing purity, $P \rightarrow 0$.

Higher-order power sums $M(m)$ with $m \ge 3$ become relevant when more than two cobosons interfere. For example, the interference of $N_2=1$ with $N_1$ cobosons reflects the normalization ratio $\chi_{N+1}/\chi_N$ \cite{Combescot2011,Combescot2003,Law2005,Chudzicki2010,Ramanathan2011}:
\eq 
P_{\text{tot}}(m)=  \frac{\chi^{\phantom \dagger}_{N_1+1} }{\chi^{\phantom \dagger}_{N_1}} P(m,0) + \left( 1- \frac{\chi^{\phantom \dagger}_{N_1+1} }{\chi^{\phantom \dagger}_{N_1}} \right) P(m,1)   \label{HOM2NNsignal} .
\en

In general, the balance between all the weights $w_0, \dots, w_{N_2}$ governs the counting statistics. Since the weights $w_p$ depend on power-sums $M(m)$ up to order $N_1+N_2$, the characteristics of the distribution $\vec \lambda$ can be established through interference signals. For ${N_1=N_2=2}$, we illustrate the decomposition (\ref{intuitivedecomposition}) in Fig.~\ref{decompositions}. The ideal boson interference pattern $P(m,0)$ is jeopardized by the finite purity $P=1/4$, the contributions of the single fermion-pair and double fermion-pair part in the wave-function lead to the altered signal $P_{\text{tot}}(m)$.

\begin{figure}[th]
\center
\includegraphics[width=\linewidth,angle=0]{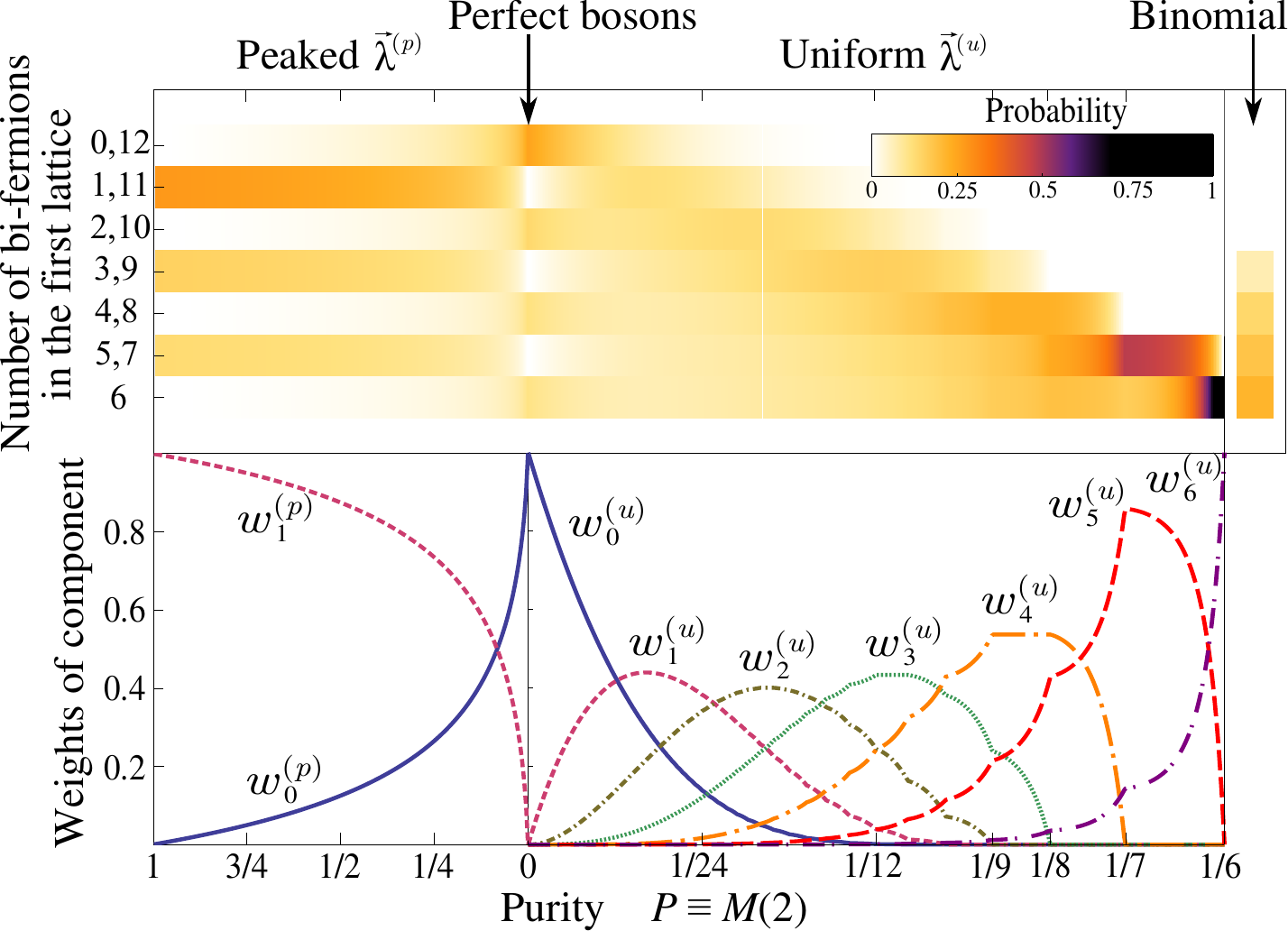} 
\caption{(color online) Upper panel: Counting statistics $P_{\text{tot}}(m)$ as a function of the purity, for the uniform $(u)$ (right-hand part) and peaked $(p)$ (left-hand part) distributions $\vec \lambda^{(u/p)}$. Lower panel: Corresponding weights $w^{(p/u)}$ of the coboson wavefunction given in (\ref{intuitivedecomposition}). We set $N_1=N_2=6$, $R=1/2$. The counting statistics is perfectly bosonic for vanishing purity, $P\rightarrow 0$, while cobosons behave as fermions for the uniform distribution and $P=1/6$. The number of non-vanishing Schmidt-coefficients in the uniform distribution is $L=\lceil 1/P \rceil$, hence the weights $w^{(u)}_l$ with $l<N-L-1$ vanish: There are at least $N_1-L-1$ pairs of fermions, which results in the kinks in the weights. The binomial distribution corresponds to the statistics of distinguishable particles. } 
\label{DynamicsCobosons}
\end{figure}

Distributions with the same purity $P$ may have different higher-order power sums $M(m)$, with consequently distinct counting statistics. Keeping $P$ constant, the counting statistics is extremized by two particular distributions: the upper bound in (\ref{boundsw0}) is saturated by peaked distribution $\vec \lambda^{(p)}$ with $\lambda^{(p)}_1 > \lambda^{(p)}_2 = \dots = \lambda^{(p)}_S$, in the limit $S\rightarrow \infty$; the lower bound is saturated by the uniform distribution $\vec \lambda^{(u)}$ with $\lambda^{(u)}_1 \le  \lambda^{(u)}_2 = \dots =  \lambda^{(u)}_{L\equiv \left \lceil 1/P \right \rceil }$, for fractional purities $P=1/L$  \cite{ourselves}. The counting statistics for $N_1=N_2=6$ is shown in Fig.~\ref{DynamicsCobosons}.   
 The weights $w_k^{u(p)}$ of the uniform (peaked) distributions differ considerably (see lower panel), which is reflected by the counting statistics (upper panel, note that $P(m)=P(12-m)$ due to symmetry). Only one Schmidt coefficient in the peaked distribution is finite in the limit $S \rightarrow \infty$, thus only the weights $w^{(p)}_0$ and $w^{(p)}_1$ are non-vanishing: the interference patterns of $12$ and of $10$ bosons take turns. Instead, all weights $w^{(u)}_{0 \le j \le 6}$ alternate for the uniform distribution. Kinks emerge at fractional values of $P$, when a new non-vanishing Schmidt coefficient emerges. For $P\rightarrow 1/6$, fully fermionic behavior is attained, and one always finds six cobosons in each lattice. 

\begin{figure}[th]
\center
\includegraphics[width=\linewidth,angle=0]{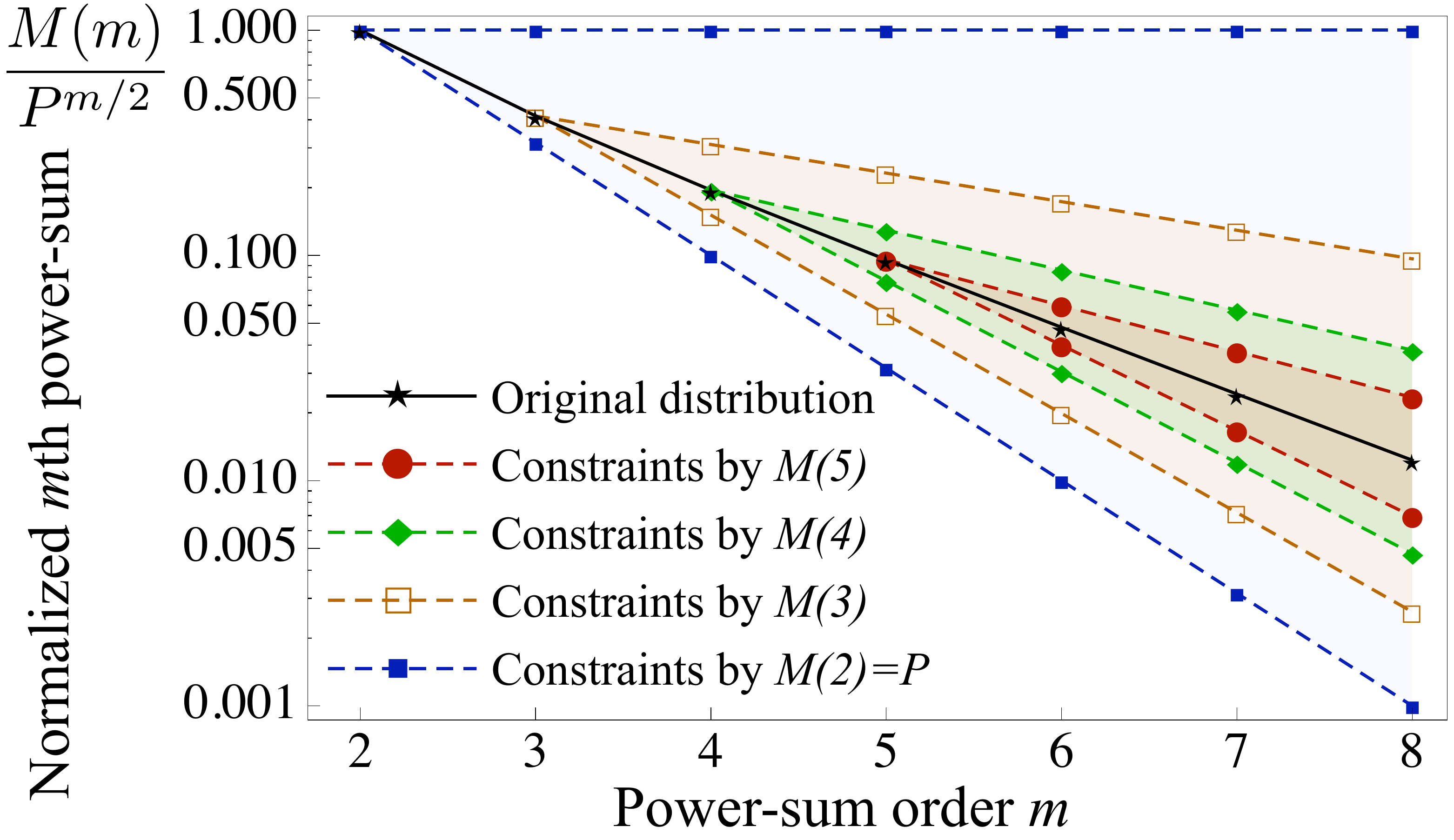} 
\caption{(color online) Normalized power-sums and constraints. The normalization to $P^{m/2}$ is chosen such that the upper bound is constant. A randomly chosen distribution $\vec \lambda$ leads to a certain hierarchy of power-sums (black stars). The measurement of interference signals with $N_1$ and $N_2$ cobosons reveals the power-sums up to order $N_1+N_2$, which leads to the indicated constraints on higher-order $M(m)$ with $m \ge N_1+N_2+1$ (blue, orange, green and red symbols), according to Eq.~(\ref{jensen}). } 
\label{constraining}
\end{figure}

The dependence of $P_{\text{tot}}(k)$ on the power-sums $M(m)$ can be used to infer the latter from measured counting statistics for different $N_1, N_2$. The purity $P$ follows immediately for $N_1=N_2=1$ via Eq.~(\ref{HOM2signal}); in general $M(m)$ is inferred by the counting statistics of a total of $N_1+N_2=m$ cobosons. Since higher-order power-sums are constrained by Jensen's and H\"older inequalities  \cite{Hardy1988},  
\eq 
 M(m-1)^{\frac{m-1 }{m-2}} \le M(m) \le M(m-1)^{\frac m {m-1}} \label{jensen} ,
\en
bounds for higher-order $M(m)$ become tighter with increasing knowledge of $M(m)$, as depicted in Fig.~\ref{constraining}. 

When the exact counting statistics cannot be retrieved and many ($N \gtrsim 1000$) cobosons are brought to interference, such as in the interference of BECs  \cite{Lucke2011a}, the granular structure of the interference pattern becomes secondary. The impact of imperfect bosonic behavior can then be incorporated into a macroscopic wavefunction approach \cite{Laloe2011}, {\it i.e.}~the number of particles is treated as the amplitude of a single-particle wavefunction. Fock-states are modeled by a random phase between the different components of the wavefunction. When the fractions $I_j=N_j/(N_1+N_2)$ of ideal bosons are prepared in the two lattices, the particle fraction $I$ in the upper lattice after beam-splitter dynamics obeys the probability distribution \cite{Laloe2011} 
\eq
{\mathcal P}_{\text{MWF}}(I;I_1,I_2) = 
\frac{1}{ \pi \sqrt{4 RT I_1I_2-(I-R I_1-T I_2)^2}} , \nonumber 
\en
for  ${4R T I_1 I_2} > {(I-R I_1-T I_2)^2}$,  while it vanishes otherwise. For cobosons, a finite fraction of fermions needs to be accounted for in each lattice. The probability distribution for the particle fraction $I$ then becomes 
\eq
{\mathcal P}(I) = 
 \int_0^{I_2} \text{d}I_f ~ \mathcal{W}(I_f) ~ {\mathcal P}_{\text{MWF}}\left(I-I_f;I_1-I_f,I_2-I_f \right) \nonumber ,
\en
where $ \mathcal{W}(I_f)$ is the probability distribution for the fraction of fermions $I_f$ in each lattice. For the uniform state $\vec \lambda^{(u)}$ with $S$ Schmidt coefficients ($P=1/S$),  
\eq 
w^{(u)}_p=\frac{N_1! N_2! (S-N_1)! (S-N_2)!}{ S! (S+p-N_1-N_2)! (N_1-p)! (N_2-p)! p!} .
 \en
The continuous limit $ \mathcal{W}^{(u)}(I_f)$ is obtained for ${N_1+N_2=:N \rightarrow \infty}$, when $N_1, N_2, p$ and $S$ are scaled linearly with $N$: 
\eq
 \mathcal{W}^{(u)}(I_f)=\lim_{N \rightarrow \infty} \left( N \cdot w^{(u)}_{(p=I_f \cdot N)} \right) = \delta \left(I_f - \rho I_1 I_2 \right) ,
\en 
and the total number of bi-fermions per Schmidt mode is constant, $\rho=N/S$. Since the number of bi-fermions in either lattice is limited by $S$, it holds $0< \rho \le 1/I_1 \le 2$. The fraction of perfect fermions is thus exactly the fraction of expected pairs of bi-fermions in the same Schmidt-mode, $\rho I_1 I_2$, which gives 
\eq 
{\mathcal P}^{(u)}(I) = {\mathcal P}_{\text{MWF}} \left(I-\rho I_1 I_2 ; I_1(1-\rho I_2), I_2(1-\rho I_1)  \right).  \nonumber 
\en 
The width $W$ of this distribution is closely related to the fraction of fermions, 
\eq 
W=4 \sqrt{ R T I_1 I_2 (1-\rho I_1) (1-\rho I_2) } ,  \hspace{0.2cm} \label{wdith} 
\en 
and becomes narrower with increasing number of bi-fermions per Schmidt mode, $\rho$. In principle, this may jeopardize Fock-state-interferometry with non-elementary particles such as neutral atoms, since the width of the intensity distribution is used to infer a small phase (which translates here to a reflectivity $R$). 

Trapped ultracold atoms typically feature very small electron-state purities of the order of $10^{-13}$ \cite{Rombouts2002,Chudzicki2010}, such that atom interferometers are not sensitive to the compositeness of the atoms. With attractively interacting fermionic atoms in tunable external potentials \cite{Serwane2011,Zurn2012}, the transition between fully bosonic (${P\rightarrow 0}$) and fully fermionic ($P \rightarrow 1$)  behavior may be implemented experimentally by varying the size of the available single-fermion space and observing the resulting interference pattern when bi-fermions are brought to interference \cite{supplementary}. 

In conclusion, even though two fermions may be arbitrarily strongly bound to a coboson with no apparent substructure,  deviations from ideal bosonic behavior can be observable in many-coboson interference.  Not the binding energy, but the entanglement between the fermions is observable on the level of the cobosons. The superposition (\ref{intuitivedecomposition}) allows to understand the partially fermionic behavior of cobosons, and ultimately leads to simple expressions for the interference of BEC (\ref{wdith}). The methods that we have exposed can be extended immediately to larger numbers of sublattices and to more complex interference scenarios \cite{tichyunpu}. 

Cobosons always constitute \emph{indistinguishable} particles; two cobosons in the two lattices share the same distribution of Schmidt coefficients $\vec \lambda$.  The impact of partial distinguishability and the effects of compositeness can actually be discriminated in the experiment: While partially distinguishable particles can be described as a superposition of perfect bosons and distinguishable particles \cite{Tichy2011,Ra2011}, cobosons exhibit the behavior of a superposition of bosons and fermions, which naturally leads to differing interference patterns in the two cases (see also the binomial distribution in Fig.~\ref{DynamicsCobosons}, which is attained for distinguishable particles).  

The role of entanglement for bosonic behavior is twofold: It circumvents the Pauli principle for composite bosons \cite{Law2005,Chudzicki2010,Combescot2003,Ramanathan2011,ourselves}, and it maintains many-particle coherence. Quantum correlations between the fermions are necessary for the bosonic exchange symmetry in the relevant parts of the wave-function that allows the representation in  Eq.~(\ref{intuitivedecomposition}). If mixed states of bi-fermions are  prepared instead of entangled states, the exchange symmetry and the encountered bosonic behavior break down -- even though  the combinatorial argument that relates to the number of accessible states remains valid. The visibility of correlation signals of, {\it e.g.}~large molecules, is thus not only affected by the mixedness of the molecules at finite temperatures, but also by the consequent loss of many-particle coherence.

\emph{Acknowledgements} This work was partially supported by the Project FQM-2445 of the Junta de Andaluc\'ia and the grant FIS2011-24540 of the Ministerio de Innovaci\'on y Ciencia, Spain. M.C.T. gratefully acknowledges support by the Alexander von Humboldt-Foundation through a Feodor Lynen Fellowship.

\clearpage 
\renewcommand{\thefigure}{\Roman{figure}}

\begin{widetext}

\setcounter{equation}{0}
\setcounter{figure}{0}
\section{Supplemental Material}

\subsection{Physical model, Hamiltonian Eq.~(1), and  preparation of the state of Eq.~(4)}
For a tangible model of composite bosons, we consider fermions of two distinguishable species, $a$ and $b$, which interact attractively via  a contact interaction $U$, and which are prepared in two weakly coupled one-dimensional lattices, as depicted in Fig.~1 of the main text. 

The single-particle tunnelling rates between the wells in horizontal (vertical) direction are denoted by  ${J_{0,h}}$ (${J_{0,v}}$), such that the Hamiltonian on the level of the individual fermions reads 
\eqq
\begin{split}
\hat H & = - U \sum_{q=1}^2 \sum_{j=1}^S \hat a^\dagger_{q,j} \hat b^\dagger_{q,j} \hat b_{q,j}^{\phantom \dagger} \hat a_{q,j}^{\phantom \dagger}   
+\frac 1 2 \sum_{q=1}^2 \sum_{j=1}^S  \epsilon_j \left( \hat a^\dagger_{q,j} \hat a_{q,j}^{\phantom \dagger} +\hat b^\dagger_{q,j}  \hat b_{q,j}^{\phantom \dagger}\right) \\
&  - \frac{ {J_{0,h}}} 2 \sum_{q=1}^2 \sum_{j=1}^{S-1} \left( \hat a^\dagger_{q,j} \hat a_{q,j+1}^{\phantom \dagger} +\hat b^\dagger_{q,j}  \hat b_{q,j+1}^{\phantom \dagger} + h.c. \right)  
 - \frac{{J_{0,v}}} 2 \sum_{j=1}^S \left( \hat a^\dagger_{1,j} \hat a_{2,j}^{\phantom \dagger} +\hat b^\dagger_{1,j}  \hat b_{2,j}^{\phantom \dagger} + h.c. \right),  
\end{split}
 \enn
 where $q=1,2$ denotes the two one-dimensional sublattices, and $j=1,\dots, S$ the potential wells along each lattice. The attractive interaction $U$ is always strong, $U \gg {J_{0,v}}, {J_{0,h}}$. Therefore, two fermions of the two different species are always bound to a bi-fermionic particle described by 
\eqq
\hat d_{q,j}^\dagger =\hat a_{q,j}^\dagger \hat b_{q,j}^\dagger ,
\enn
which fulfills the algebraic properties of a hardcore-boson operator \cite{Girardeau}, 
\eqq
\begin{split}
p\neq q \text{ or } j\neq m : [ \hat d_{p,j}^\dagger , \hat d_{q,m}^\dagger ]=0, \\
\left( \hat d_{q,j}^\dagger \right)^2=\left( \hat d_{q,j} \right)^2= 0 , \
\{ \hat d_{q,j}^\dagger , \hat d_{q,j}^\dagger \}=1 , 
\end{split}
\enn 
under the assumption that the numbers of fermions of each species coincide.

Due to the strong attractive interaction between the fermions, they can only tunnel as a pair through off-resonant processes, with rates 
\eqq {J_{h}}=\frac{{J_{0,h}}^2}{U} , \ \ {J_{v}}=\frac{{J_{0,v}}^2}{U},   \enn
 for the interlattice and intralattice directions, respectively. We thus recover the effective Hamiltonian given by Eq.~(1) in the main text:
\eqq
 \hat H  =  
 \left[  - \frac {J_h} 2  \sum_{q=1}^2 \sum_{j=1}^{S-1}  \hat d^\dagger_{q,j} \hat d_{q,j+1}^{\phantom \dagger }   
 - \frac {J_v} 2 \sum_{j=1}^S  \hat d^\dagger_{1,j} \hat d_{2,j}^{\phantom \dagger }    + h.c.  \right] 
 +\sum_{q=1}^2 \sum_{j=1}^S  \epsilon_j \left( \hat d^\dagger_{q,j} \hat d_{q,j}^{\phantom \dagger }  \right) . \label{hamil}  \enn

While the main purpose of our analysis is an investigation of compositeness for the collective interference exhibited by the state Eq.~(4) in the main text, it is worthwhile to indicate a procedure by which such a state with many cobosons in the same state may be produced as the result of an experimental protocol. 

We may start with an extended lattice with a total of $S N_1$ sites (remember $N_1 \ge N_2$ by assumption), such that site $k S$ (with $k=1, \dots, N_1-1$) is coupled to site $k S+1$ with a strength $J_s \ll J_h$, and $\epsilon_{k S+ j}=\epsilon_{j}$ for $k=1,\dots , N_1$. 
We exploit the exact mapping of hardcore-bosons to fermions in one dimension to obtain the ground-state of $N_1$ and $N_2$ bi-fermions in the first and second extended lattice as a direct product of the lowest $N_1$  and $N_2$ single-particle states, respectively,
\eqq
\ket{\text{GS}(N_1,N_2)}= \left[ \prod_{j=1}^{N_1} \left( \sum_{l=1}^{S N_1} \omega_{j,l} \hat d^\dagger_{1,l}   \right)  \right] \left[ \prod_{j=1}^{N_2} \left( \sum_{l=1}^{S N_1} \omega_{j,l} \hat d^\dagger_{2,l}   \right) \right]  \ket{0} ,
\enn
where the matrix $\omega_{j,l}$ contains the coefficients $l=1\dots S N_1$ of the $j$th single-particle eigenfunction in one sublattice. After preparing the ground state with $N_1$ and $N_2$ particles in each sublattice, we project away the component of the many-body wavefunction in which particles are present in the sites $S+1\dots S N_1$ of either lattice.  Consequently, a particle initially prepared in the $j$th eigenstate in lattice $q$ is projected -- with a finite probability -- onto the state created by 
\eqq 
\sum_{l=1}^{S} \omega_{j,l} \hat d^\dagger_{q,l}  .
\enn
Our above choice of the potential landscape with $J_s \ll J_h$ ensures that the  $N_1$ energetically lowest single-particle wave-functions have a node around $l=S$ and that the coefficients $\omega_{j,l}$ with $1\le l \le S$ are very similar for the first $N_1$ eigenfunctions, 
\eqq
 2 \le j \le N_1 ,  1 \le l \le S : \omega_{j,l} \approx \omega_{1,l} .
\enn

 After projecting out components that populate the auxiliary sites $j=S+1 \dots S N_1$, the many-particle-state is consequently very close to a state of $N_q$-fold population of the co-boson state
\eqq 
\hat c_{q}^\dagger = \alpha \sum_{l=1}^{S} \omega_{1,l} \hat d^\dagger_{q,l} = \sum_{l=1}^{S} \sqrt{\lambda_l} \hat d^\dagger_{q,l} , \label{cobosoncreation}
\enn
where $\alpha$ ensures normalization and we thus set $\lambda_l =  |\alpha \omega_{1,l}|^2$. Note that the Pauli principle ensures that unlike for hardcore bosons the preparation of fermions into this state is impossible: One then always finds at least one particle in the sites $S+1\dots S N_1$, and the projection always fails. 

In the main text, we take the initial state (4) as given, and thus consider coboson operators for the two lattices $q=1,2$ as given by Eq.~(\ref{cobosoncreation}). 
By appropriately modelling the local energies $\epsilon_j$, the resulting distribution of Schmidt coefficients $\lambda_l$ can be modelled to a very wide extent.

\subsection{Behavior of cobosons under beam-splitter dynamics, Eqs.~(7,8)}
We would like to describe the counting statistics of the state $\ket{\Psi}$ of Eq.~(4) in the main text in an efficient manner. For that purpose, we insert the definition of the coboson creation operator, Eq.~(\ref{cobosoncreation}), into the initial state, Eq.~(4):
\eqq 
\left[ \hat c_1^\dagger \right]^{N_1} \left[ \hat c_2^\dagger \right]^{N_2} =  \label{thebsoum}     \sum_{ \substack{ k_1 \neq \dots \neq k_{N_1}  \\ l_1 \neq \dots \neq l_{N_2}} }^{1 \le k_j , l_j  \le S} 
\left[  \prod_{m=1}^{N_1} \sqrt{\lambda_{k_m}} \hat d^\dagger_{1,k_m} \right]
 \left[  \prod_{n=1}^{N_2} \sqrt{\lambda_{l_n}} \hat d^\dagger_{2,l_n} \right]  .
\enn
All indices $k_m$ ($l_n$) appertain to the upper  (lower) lattice. It may occur that $k_m=l_n$ for some $m,n$, \emph{i.e.}~two bi-fermions can occupy the same well $j$ in the two different lattices. The sum (\ref{thebsoum}) can be written in terms with a given number of pairs of indices $p$ that fulfill ${k_m=l_n}$. There can be between none and $N_2$ of such pairs (remember $N_1 \ge N_2$): In the former case, $k_m \neq l_n$ for all $m$, $n$; in the latter, $k_m=l_m$ for all $m \le N_2$ (disregarding permutation of indices). The state initial state $\ket{\Psi}$ thus becomes 
  \eqq 
\ket{\Psi}=\sum_{p=0}^{N_2} \ket{\Phi(p)} , \label{formaldecomposition}
\enn
 where
\eqq
\ket{\Phi(p)} :=\sqrt{ {N_2 \choose p}  {N_1 \choose p} \frac{p! }{\chi_{N_1}\cdot N_1! \cdot \chi_{N_2} \cdot N_2! } }  \label{phidef} 
\sum^{(N_2,p)}  \left[ \prod_{m=1}^{N_1}   \sqrt{\lambda_{k_m}} \hat d^\dagger_{1,k_m} \right] \left[  \prod_{n=1}^{N_2} \sqrt{\lambda_{l_n} }  \hat d^\dagger_{2,l_n} \right] \ket{0} ,
 \enn
and the sum $ \sum^{(N_2,p)}$ runs over all indices $k_m$, $l_n$ ($1 \le m \le N_1, 1 \le n \le N_2$) that fulfill 
\eqq
\begin{split}  i \neq j &: k_i \neq k_j, l_i \neq l_j  ,\\
 \ 1 \le m \le p &: k_m= l_m , \\
  \  p < m \le N_1, p< n \le N_2 &: k_m \neq l_n . \end{split} \enn 
Reordering indices and setting $\tilde N := N_1+N_2-2p$, we can rewrite the sum (\ref{phidef}) as
\eqq \label{rewrittenEquation}
\begin{split}
\ket{\Phi(p)} = &\sqrt{ {N_2 \choose p}  {N_1 \choose p} \frac{p! }{\chi_{N_1}\cdot N_1! \cdot \chi_{N_2} \cdot N_2! } }  \times \\
&
\sum_{l_1\neq l_2 \neq \dots \neq l_p}^{1 \le l_j \le S } 
   \sum_{ \substack{ 1 \le r_{1} < \dots < r_{\tilde N} \le S }}^{ \forall l,m: r_l \neq l_m }
 \sum_{\sigma \in S_{\{ r_1, \dots , r_{\tilde N } \}} } 
 \underbrace{ \left[ \prod_{j=1}^{p } \lambda_{l_j}    \hat  d^\dagger_{1,{l_j} }   \hat  d^\dagger_{2,{l_j} } \right] }_{\text{fermionic}}
 \left[ \prod_{j=1}^{\tilde N } \sqrt{\lambda_{\sigma(j)} }  \right]
 \underbrace{  \left[ \prod_{m=1}^{N_1-p}   \hat  d^\dagger_{1,\sigma(m)} \right]
 \left[  \prod_{n=1}^{N_2-p}  \hat  d^\dagger_{2,\sigma(N_1-p+n)} \right]  }_{\text{bosonic}},
 \end{split}
\enn
where the indices $r_1\dots r_{\tilde N}$ replace the $k_{m>p}$ and $l_{m>p}$ and $S_{\{r_1, \dots , r_{\tilde N} \}}$ denotes the permutations of the  $r_j$. 

By inspecting (\ref{rewrittenEquation}), we can now infer the time-evolution of each $\ket{\Phi(p)}$ component of the many-coboson state (\ref{formaldecomposition}), as it is induced by the Hamiltonian (1) in the here-considered limit $J_h \ll J_v$: Each summand in (\ref{rewrittenEquation}) contains $p$ pairs of bi-fermions that occupy the same well in the two lattices -- they take into account the summands with $k_m=l_m$ ($1 \le m \le p$) of  Eq.~(\ref{phidef}). Due to the Pauli principle, these bi-fermions cannot tunnel and thus behave fermionic. The other  bi-fermions described by the indices $r_{1}, \dots, r_{\tilde N}$ (which correspond to the indices $n,m>p$ in Eq.~(\ref{phidef})) always occupy different wells, such that the Pauli principle does not apply and tunneling is possible.  One such summand of (\ref{rewrittenEquation}) is  depicted in Fig.~\ref{SumRepresentation.pdf}.  
Since the state of the bi-fermions that can tunnel is fully symmetric under the exchange of any two bi-fermions between the lattices (note the sum over all permutations of the lattice indices in (\ref{rewrittenEquation})), it is manifestly bosonic, as also illustrated in Fig.~\ref{InterferencePicture.pdf}.

\begin{figure}[th]
\center
\includegraphics[width=.5\linewidth,angle=0]{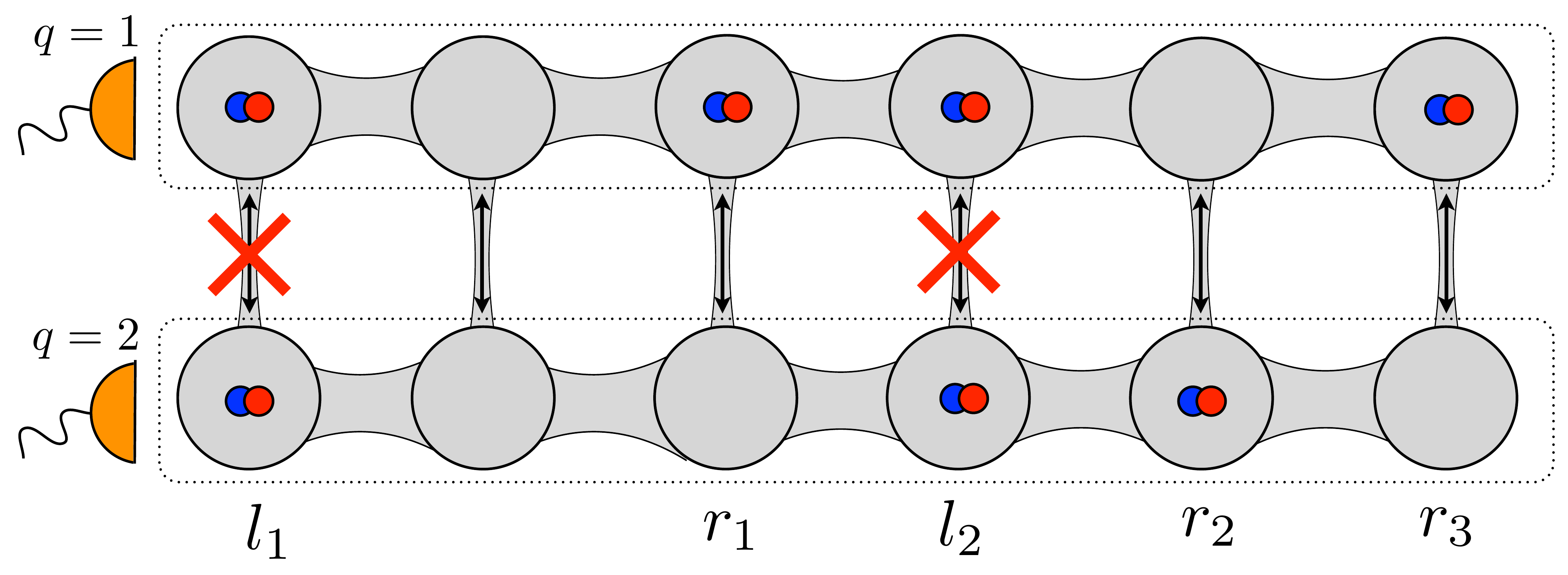} 
\caption{One summand of Eq.~(\ref{rewrittenEquation}), with $N_1=4, N_2=3$: Two pairs of bi-fermions $(p=2$) are located in the same wells $l_1$ and $l_2$, and cannot tunnel. All other bi-fermions can tunnel, and interfere with other terms in the sum (see Fig.~\ref{InterferencePicture.pdf}). } 
 \label{SumRepresentation.pdf}
\end{figure}

\begin{figure}[th]
\center
\includegraphics[width=.6\linewidth,angle=0]{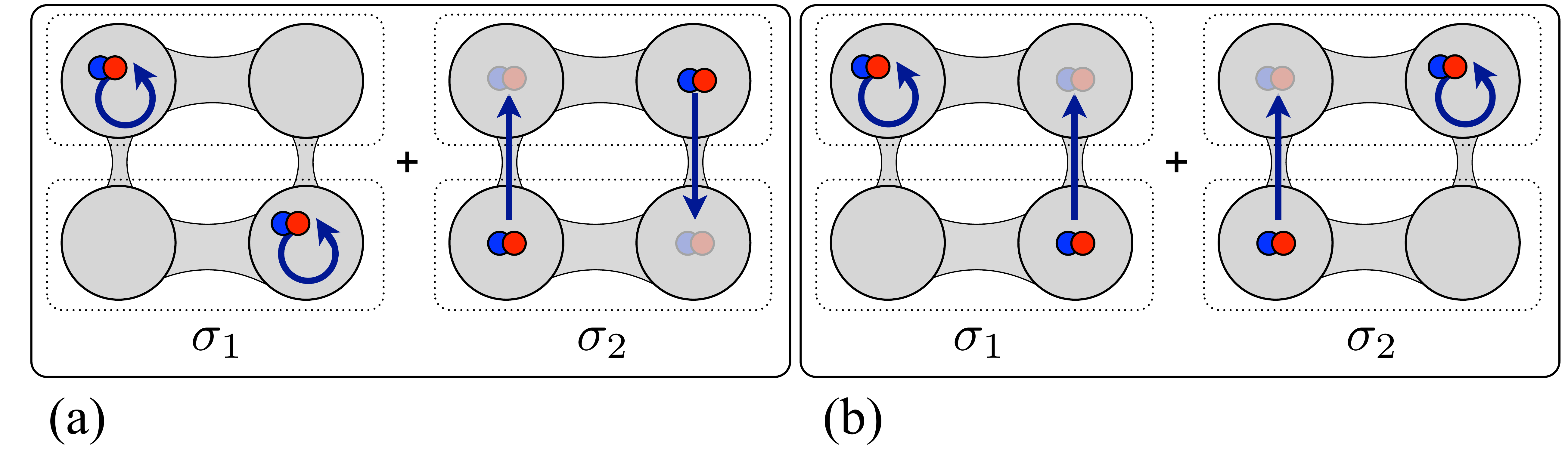} 
\caption{Emergence of Hong-Ou-Mandel-like bunching for bi-fermions. Each component of the wave-function in (\ref{rewrittenEquation}), corresponding to a permutation $\sigma_1$, interferes with another component $\sigma_2$ in which the bi-fermions swap the wells they occupy. (a) Destructive interference: The two  processes (both bi-fermions remain in the same lattice, or both bi-fermions tunnel) lead to the same final state with one particle in each lattice. The right-hand-side process, however, acquires a phase of $i^2=-1$ due to tunneling, such that the two processes interfere destructively. (b) Constructive interference: The final state with both bi-fermions in the same lattice is fed by two processes with one tunnelling event, i.e. both paths acquire the same phase and constructive interference takes place. } 
 \label{InterferencePicture.pdf}
\end{figure}

The total state $\ket{\Phi(p)}$ thus exhibits the same counting statistics as a state $\ket{\phi(p)}$ of $p$ pairs of distinct fermions  and $N_1+N_2-2p$  perfect bosons (Eq.~(9) in the main text): 
\eqq
\ket{\phi(p)} =
 \left[ \prod_{q=1}^2  \frac{\left( \hat g_q^\dagger \right)^{N_q-p} }{\sqrt{(N_q-p)! }} \right]
 \left[ \prod_{j=1}^p \hat f_{1,j}^\dagger \hat f_{2,j}^\dagger \right] \ket{0} , \label{Phipstate}
\enn
where $\hat g_{q}^\dagger$ ($\hat f_{q,j}^\dagger$) creates a boson ($j$-type fermion) in the sublattice $q$ while the actual location along the lattice is  omitted (remember that the number of bi-fermions in each lattice is counted, independently of the location of the bi-fermions along the lattice). Since no interference between different $p$ occurs, the initial state $\ket{\Psi}$ (Eq.~(4)) behaves like a superposition of states $\ket{\phi(p)}$ with different numbers of ideal fermions, in analogy to partially distinguishable particles \cite{Tichy2011}: 
\eqq
\ket{\psi}= \sum_{p=0}^{N_2} \sqrt{w_p} \ket{\phi(p)} , \label{substitute}
\enn
where 
\eqq
\label{weight}
w_p= \braket{\Phi(p)}{\Phi(p)}  = {N_1 \choose p }  {N_2 \choose p } ~\frac{p!}{\chi_{N_1} \chi_{N_2} } ~\Omega(\{ \underbrace{2,\dots, 2}_p,\underbrace{1,\dots ,1}_{N_1+N_2-2p}\}) ,
\enn
is the weight of the component with $p$ pairs of fermionically behaving bi-fermions, and $\Omega(\{ x_1, \dots x_N \})$ is given by Eq.~(5) in the main text. 

\subsection{Evaluation of counting statistics, Eq.~(11)}
The counting statistics that is exhibited by the substitute state $\ket{\psi}$, Eq.~(\ref{substitute}), can be inferred by inserting the single-particle time-evolution for the creation operators for each state $\ket{\phi(p)}$, 
\eqq
\begin{split} 
\hat g^\dagger_q &  \rightarrow  i \sqrt{R} \hat g^\dagger_q +  \sqrt{T} \hat g^\dagger_{3-q}  \\
\hat f^\dagger_{q,j} &  \rightarrow  i \sqrt{R} \hat f^\dagger_{q,j}+ \sqrt{T} \hat f^\dagger_{3-q,j}  , 
\end{split}
\enn
where $R=\cos^2(J_v t/2)$ and $T=\sin^2(J_v t/2)$ are the reflection and transmission coefficients of the beam-splitter dynamics. 

The probability $P(m,p)$ to find $m$ particles in the upper lattice can then be inferred by taking the overlap of the state $\ket{\phi(p)}$ after time-evolution with the Fock-state of $m-p$ bosons and $p$ fermions in the first mode, 
\eqq 
 \frac{\left(\hat g_1^\dagger \right)^{m-p} \left( \hat g_2^\dagger\right)^{N-m-p}}{\sqrt{ (m-p)! (N-m-p)!}} \prod_{j=1}^p f_{1,j}^\dagger f_{2,j}^\dagger   \ket{0} .
\enn
The evaluation of the overlap can be done following the methods presented in Refs.~\cite{Laloe2011,tichyunpu,Tichy2011}.

\end{widetext}

\end{document}